\documentclass[aps,prl,twocolumn,superscriptaddress]{revtex4}

\usepackage{graphicx}         % needed for figures
\usepackage{bm}               % for math
\usepackage{amssymb}          % for math
\usepackage{amsmath}          % for aligned
\usepackage{CJK}              % for Chinese
\usepackage[dvipsnames]{xcolor} % for color text
\usepackage{orcidlink}

\newcommand {\eV}          {\,\rm eV}
\newcommand {\pc}          {\,\rm pc}
\newcommand {\kpc}         {\,\rm kpc}

\newcommand {\kyr}         {\,\rm kyr}
\newcommand {\Msun}        {\,M_{\odot}}
\newcommand {\MBH}         {M_{\rm BH}}
\newcommand {\MdotBH}      {\dot{M}_{\rm BH}}
\newcommand {\MdotBondi}   {\dot{M}_{\rm Bon}}
\newcommand {\MdotBondiOri}{\dot{M}_{\rm Bon0}}
\newcommand {\MdotEdd}     {\dot{M}_{\rm Edd}}
\newcommand {\tBondi}      {t_{\rm Bon}}
\newcommand {\rBondi}      {r_{\rm Bon}}

\newcommand {\fref}[1]     {Fig.~\ref{#1}}
\newcommand {\tref}[1]     {Table~\ref{#1}}

\newcommand {\be}          {\begin{equation}}
\newcommand {\ee}          {\end{equation}}

\begin{document}
\begin{CJK*}{UTF8}{bkai}

\title{Boosting Supermassive Black Hole Growth in the Early Universe by Fuzzy Dark Matter Solitons}

%authors and affiliations
\author{H.-H. Sandy Chiu (邱懷萱)\orcidlink{0000-0002-7401-382X}}
\affiliation{Department of Physics, National Taiwan University, Taipei 10617, Taiwan}
\affiliation{Department of Astronomy, University of Michigan, Ann Arbor, MI 48109-1107, USA}

\author{Hsi-Yu Schive (薛熙于)\orcidlink{0000-0002-1249-279X}}
\email{hyschive@phys.ntu.edu.tw}
\affiliation{Department of Physics, National Taiwan University, Taipei 10617, Taiwan}
\affiliation{Institute of Astrophysics, National Taiwan University, Taipei 10617, Taiwan}
\affiliation{Center for Theoretical Physics, National Taiwan University, Taipei 10617, Taiwan}
\affiliation{Physics Division, National Center for Theoretical Sciences, Taipei 10617, Taiwan}

\author{Hsiang-Yi Karen Yang (楊湘怡)\orcidlink{/0000-0003-3269-4660}}
\affiliation{Physics Division, National Center for Theoretical Sciences, Taipei 10617, Taiwan}
\affiliation{Institute of Astronomy and Department of Physics, National Tsing Hua University, Hsinchu 30013, Taiwan}

\author{Hsinhao Huang (黃新豪)\orcidlink{/0000-0002-7368-1324}}
\affiliation{Department of Physics, National Taiwan University, Taipei 10617, Taiwan}

\author{Massimo Gaspari\orcidlink{0000-0003-2754-9258}}
\affiliation{Department of Physics, Informatics and Mathematics, University of Modena and Reggio Emilia, 41125 Modena, Italy}

\date{\today}

% abstract
% ----------------------------------------------
\begin{abstract}
Observations of massive supermassive black holes (SMBHs) in the early universe challenge existing black hole formation models. We propose that soliton cores in fuzzy dark matter (FDM) offer a potential solution to this timing problem. Our FDM cosmological zoom-in simulations confirm that for a particle mass $m_{\rm FDM}\sim 10^{-22}\eV$, solitons are well developed at redshift $z \sim 7$ with masses of $\sim10^9\Msun$, comparable to the observed SMBHs. We then demonstrate using hydrodynamic simulations that, compared to cold dark matter, these high-$z$ massive FDM solitons with mass $M_s$ can provide additional gravitational potential to accrete gas and boost the Bondi accretion rate of a growing black hole seed with mass $\MBH$ by up to two to four orders of magnitude, in the regime of efficient cooling and negligible radiation pressure. This accretion boosting mechanism is effective for $10^{-22}\eV \lesssim m_{\rm FDM} \lesssim 10^{-20}\eV$ and potentially beyond as long as $M_s > \MBH$.
\end{abstract}

\maketitle
\end{CJK*}

% introduction
% ----------------------------------------------
\textit{Introduction}.---Observations of supermassive black holes (SMBHs) at high redshifts $z \gtrsim 7$ \cite{banados_800-million-solar-mass_2018, yang_poniuena_2020, larson_ceers_2023, kokorev_uncover_2023, kovacs_candidate_2024, bogdan_evidence_2024, bosman_mature_2024, suh_super-eddington-accreting_2024}
% z>8 + M=1e7-1e8 Msun reference: larson_ceers_2023, kokorev_uncover_2023, kovacs_candidate_2024
with masses of $\MBH \gtrsim 10^7$--$10^9\Msun$ represent a significant challenge to existing black hole (BH) formation models \cite{valiante_formation_2017, woods_titans_2019, inayoshi_assembly_2020, zhu_formation_2022}. These SMBHs may originate from various BH seeds, including massive seeds exceeding $10^4 \Msun$ formed through direct gas collapse \cite{wise_formation_2019, sassano_light_2021, mayer_direct_2024}, lighter seeds ($\sim 10^2$--$10^3 \Msun$) formed from Population III stars \cite{bromm_formation_2013, greif_numerical_2015, sugimura_birth_2020}, or intermediate-mass seeds ($\sim 10^3$--$10^4 \Msun$) formed from runaway collisions in dense star clusters \cite{omukai_can_2008, devecchi_formation_2009}.
To reach the observed SMBH mass at such high redshifts, cosmological simulations indicate that these BH seeds must undergo rapid accretion at rates exceeding those predicted by standard accretion models. The accretion of heavy seeds must exceed the Bondi accretion rate \cite{jeon_observability_2023, jeon_physical_2024}, which characterizes the steady-state, spherically-symmetric, and adiabatic accretion onto a point mass from a uniform gas \cite{bondi_spherically_1952}\footnote{Although the applicability of Bondi accretion in realistic astrophysical environments is questionable due to the turbulence and heating/cooling processes in diffuse halos \cite{gaspari_2020}, we primarily use it here as a reference for calculating the boosting factor.}. In contrast, lighter seeds need to accrete at or above the Eddington limit \cite{smole_smbh_2015, inayoshi_hyper-eddington_2016, pezzulli_sustainable_2017, mayer_super-eddington_2019, inayoshi_assembly_2020, maiolino_small_2024, shi_seed_2024} for extended periods. Maintaining such high accretion rates is challenging because the Bondi accretion rate can only exceed the Eddington limit in extremely dense environments \cite{2012MNRAS.425.2892W}, and feedback from the active galactic nucleus (AGN) further alters the mass accretion \citep{wurster_comparative_2013,gaspari_2020,valentini_impact_2020, yao_active_2021}.

Ultra-light dark matter (ULDM) \cite{Peebles2000, Guzman2000, Goodman2000, Bohmer2007, Sikivie2009, Marsh2016, hui_ultralight_2017, hui_wave_2021, ChadhaDay2022}, also known as scalar field dark matter or wave dark matter, is one of the leading dark matter candidates. It consists of ultra-light bosons with a broad particle mass range of $\sim 10^{-22}$--$10^{-6}\eV$ and is well motivated by quantum chromodynamics axions \cite{Wilczek1978} and string theory \cite{Arvanitaki2010}. Extensive studies have examined ULDM through various astrophysical probes \cite{Irsic2017, Rogers2021, Davoudiasl2019, Brito2020, Hui2023, Chan2020, Amruth2023, Marsh2019, Schive2020, Hayashi2021, Tsai2023, An2024, Meyer2020, Ajello2016, DeMartino2017, Blas2017, Smarra2023, Manzari2024, GomezBanon2024, Beadle2024} and ground-based experiments \cite{VanTilburg2015, Antypas2019, Stadnik2023, Bloch2023, Savalle2021, Terrano2019, Fierlinger2024, Sikivie2021, Centers2019}. The high sensitivity of gravitational-wave detectors also offers a promising detection method \cite{Branca2017, Grote2019, Vermeulen2021, Gottel2024, Duque2024}. Furthermore, heating from the decay of axion-like particles may inhibit molecular hydrogen formation, potentially leading to the direct collapse of gas clouds into SMBHs \cite{Lu2024}.

ULDM in the mass range of $m_{\rm FDM}\sim 10^{-22}$--$10^{-20}\eV$ is also referred to as fuzzy dark matter (FDM) \cite{hu_fuzzy_2000}, characterized by distinctive wave-like features on sub-kiloparsec scales, such as density granulation and the suppression of low-mass halos \cite{niemeyer_small-scale_2020}. It may address small-scale challenges associated with cold dark matter (CDM) \cite{Spergel2000, Weinberg2015}, which remain a topic of debate (e.g., \cite{Kim2018}).
In particular, while maintaining the same large-scale structure as CDM, FDM predicts a unique structure at the center of each dark matter halo --- a soliton core, which is a stationary, ground-state solution of the Schr\"{o}dinger-Poisson equation \cite{schive_cosmic_2014, schive_understanding_2014, mocz_first_2019}. The soliton half-density radius $r_s$ and total mass $M_s$ as a function of the host halo mass $M_h$ and redshift $z$, derived from both numerical simulations \cite{schive_cosmic_2014, schive_understanding_2014, chen_jeans_2017, Chan2022} and analytical models assuming thermal equilibrium between halos and solitons \cite{schive_understanding_2014, Bar2019, Kawai2024}, are
\begin{equation}
    r_s = 0.135\ m_{22}^{-1}  \left(\frac{\zeta'(z)}{\zeta'(7)}\right)^{-1} \left(\frac{M_h}{10^{11} \Msun}\right)^{-1/3}  \kpc
\label{eq:rs}
\end{equation}
and 
\begin{equation}
    M_s = 1.68\times 10^9 \ m_{22}^{-1} \left(\frac{\zeta'(z)}{\zeta'(7)}\right) \left(\frac{M_h}{10^{11} \Msun}\right)^{1/3} \Msun,
\label{eq:Ms}
\end{equation}
where $m_{22}\equiv m_{\rm FDM}/10^{-22}\eV$, $\zeta'=(1+z)^{1/2}\zeta(z)^{1/6}$ with $\zeta(z)=(18\pi^2+82(\Omega_m(z)-1)-39(\Omega_m(z)-1)^2)/\Omega_m(z)\sim 180$ at $z\ge 1$ \cite{Bryan1998}, and $\Omega_m$ is the matter density parameter. $M_s$ is roughly 4.2 times larger than the mass enclosed within $r_s$. Importantly, it remains uncertain whether Eqs. (\ref{eq:rs}) and (\ref{eq:Ms}) can be reliably applied to massive halos at high redshifts, necessitating the FDM cosmological simulations performed in this study.

In this Letter, we first conduct FDM cosmological zoom-in simulations with $m_{22} \sim 1$ to confirm that FDM halos with $M_h \sim 10^{11}\Msun$ at $z \sim 7$ have well-developed solitons with masses of $\sim10^9\Msun$, comparable to SMBHs and in line with Eqs. (\ref{eq:rs}) and (\ref{eq:Ms}). We then demonstrate using hydrodynamic simulations that, compared to CDM, these high-$z$ massive FDM solitons can provide additional gravitational potential to accrete gas and boost the BH mass accretion rate $\MdotBH$ by up to two to four orders of magnitude when cooling is efficient and radiation pressure is ignored. This additional boost provided by FDM solitons offers a potential explanation for the rapid growth of SMBHs in the early universe, with accretion rates significantly exceeding those predicted by the conventional Bondi model.

% FDM simulations
% ----------------------------------------------
\textit{High-z FDM solitons}.---We conduct FDM cosmological zoom-in simulations targeting a halo with $M_h \sim 7\times10^{10} \Msun$ at $z=6.87$ using the GPU-accelerated adaptive refinement code \texttt{GAMER} \cite{schive_gamer-2_2018}. We investigate $m_{22}=1.6$ and $0.8$ in a cubic domain with a comoving side length of 15 Mpc, a $512^3$ root grid, and up to nine refinement levels, achieving a maximum comoving resolution of 57 pc. Both the simulation domain length and maximum resolution are about an order of magnitude greater than those in \cite{mocz_first_2019}. The code employs a new hybrid scheme integrating a fluid formulation on coarser grids for large scales and a wave formulation on refined grids for small scales \cite{Kunkel2024} (see also \cite{Schwabe2022}). We construct the initial conditions at $z=100$ using \texttt{AxionCAMB} \cite{AxionCAMB} and \texttt{MUSIC} \cite{MUSIC}. For comparison, we also conduct CDM simulations using the code \texttt{GADGET-2} \cite{Gadget2} with $1024^3$ particles and with a similar initial condition, except that there is no small-scale suppression by FDM quantum pressure in the initial power spectrum.

\begin{figure*}[ht!]
\includegraphics[width=\textwidth]{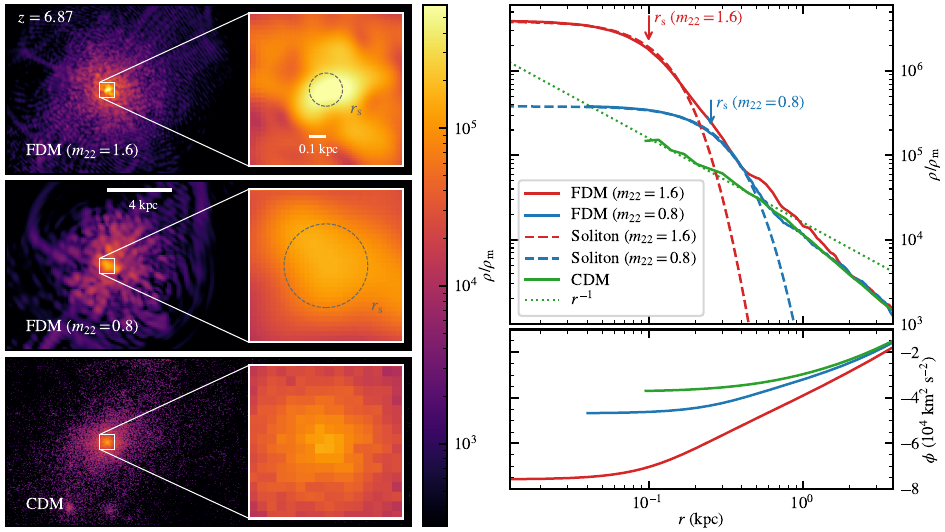}
\caption{
FDM cosmological zoom-in simulations. (Left) Projected density in a $1 \kpc$ thick slab through a $\sim 7\times10^{10} \Msun$ halo with $m_{22}=1.6$ (top) and $0.8$ (middle) at $z=6.87$, normalized to the mean matter density $\rho_{m}$; a CDM counterpart is shown for comparison (bottom). Insets highlight the central regions, with dashed circles indicating the soliton radii. (Right) Density and gravitational potential profiles of the FDM and CDM halos (solid lines) and soliton solutions of \cite{schive_cosmic_2014} (dashed lines). Compared to CDM, distinct and massive soliton cores with $r_s \sim 0.1\kpc$ and $M_s \sim 10^9 \Msun$ form in FDM, in agreement with Eqs. (\ref{eq:rs}) and (\ref{eq:Ms}), providing additional gravitational potential to accrete gas and boost SMBH growth in the early universe (see Figs. \ref{fig:profile}--\ref{fig:grid}).
}
\label{fig:zoom-in-simulation}
\end{figure*}

\fref{fig:zoom-in-simulation} shows the density and gravitational potential distributions of this target halo. FDM halos with $m_{22}=0.8~(1.6)$ formed at $z \sim 9~(11)$ and had well-developed solitons by $z \sim 7$ with $r_s \sim 0.25~ (0.10) \kpc$ and $M_s \sim 1.4\times10^9~(8.7\times10^8) \Msun$, consistent with Eqs. (\ref{eq:rs}) and (\ref{eq:Ms}). The larger the $m_{22}$, the earlier the halos and solitons form. These solitons match well the analytical soliton profile \cite{schive_cosmic_2014} and result in a significantly deeper central gravitational potential than the CDM halo. Additionally, our target halos, with $M_h \sim 7\times10^{10} \Msun$ at $z \sim 7$, are common in both CDM and FDM with $m_{22} \gtrsim 1$, as this mass scale exceeds the FDM half-mode mass, $M_{1/2} = 3.8\times10^{10} m_{22}^{-4/3} \Msun$, below which the FDM halo mass function drops significantly compared to CDM \cite{Schive2016}. These validations reinforce the soliton setup in the following hydrodynamic simulations for investigating how this additional gravitational potential of solitons may boost gas accretion and SMBH growth in the early universe.

% fig 1
% ----------------------------------------------

% BH-soliton accretion setup
% ----------------------------------------------
\textit{BH-soliton accretion setup}.---After confirming the existence of high-z solitons, we examine how the soliton gravity boosts the accretion rate of a growing BH seed.
We perform hydrodynamic simulations with \texttt{GAMER} to investigate mass accretion from a uniform gas initially at rest onto a central BH embedded in an external soliton potential $\phi_{s0}$ \cite{chen_jeans_2017}. To measure $\MdotBH$, we adopt the well-tested prescription in \cite{gaspari_chaotic_2013}, including a spherical `void' region to model the BH horizon, within which we reset the density, temperature, and velocity to negligible values after each time-step $\Delta t$, and infer $\MdotBH$ from the mass difference over $\Delta t$. In order to find the steady-state solution, we fix $\MBH$ and perform simulations until $\MdotBH$ saturates ($\sim$\,1000 Bondi times). The gas self-gravity is ignored as it is much weaker than the BH and soliton gravity.

To minimize numerical errors associated with the rapid accretion caused by a massive soliton, we gradually introduce the soliton potential by $\phi_{s}(r,t) = \phi_{s0}(r)\tanh(t/\beta t_g)$, where $t_g=GM_s/4.2c_s^3$, $G$ is the gravitational constant, and $c_s$ is the ambient sound speed. $\beta \sim 0.1$--$10$ is a dimensionless parameter to ensure the mass accretion remains adiabatic during the entire simulation. 

Table \ref{tab:param} lists the adopted simulation parameters. We use an ideal gas equation of state with a range of effective adiabatic index $\gamma=1$--$5/3$, mimicking the composite effect of non-adiabatic micro-scale processes, which are often dominated by radiative cooling in astrophysical halos. We probe the ambient temperature $T_0=1$--$10\eV\sim10^4$--$10^5$\,K to account for different primordial environments \cite{park_radiation-driven_2017,palous_can_2020,park_accelerated_2022,inayoshi_rapid_2022}. To investigate how the soliton gravity enhances the accretion rate of a growing BH seed, we fix a typical seed of $\MBH=10^5\Msun$ and probe $m_{22}=1$--$10$ to ensure $M_s \gg \MBH$. Note that within this parameter space, $r_s \gg \rBondi$, indicating that the soliton and BH gravity dominate in the outer and inner regions, respectively.

% table 1
% ----------------------------------------------
\begin{table}[t!]
\caption{Simulation parameters. Parameters in the lower table are derived from those in the upper table, assuming $\gamma=1.2$. Derived parameters are presented in ascending order of $T_0=1$--$10\eV$ for Bondi-related parameters and $m_{22}=1$--$10$ for soliton parameters.}
\label{tab:param}
\begin{tabular}{ll}
Parameter                    & Value                 \\ \hline
BH mass ($\MBH$)             & $10^5 \Msun$          \\
Adiabatic index ($\gamma$)   & 1--5/3                \\
Ambient density ($\rho_0$)   & $10^{-23}\,\rm{g/cm}^3$ \\
Atomic mass fraction         & 0.62                  \\
Ambient temperature ($T_0$)  & 1--10$\eV$            \\
FDM particle mass ($m_{22}$) & 1--10                 \\
Halo mass ($M_h$)            & $10^{11} \Msun$     \\
Redshift ($z$)               & 7                     \\ \hline
Bondi radius ($\rBondi$)     & 2.3--0.23$\pc$       \\
Bondi time ($\tBondi$)       & 165--5.2$\kyr$       \\
Bondi sonic point            & 0.81--0.081$\pc$     \\ 
Soliton radius ($r_s$)       & 135--13.5$\pc$       \\
Soliton mass ($M_s$)         & 16.8--1.68$\times 10^8\Msun$
\end{tabular}
\end{table}

We adopt a cubic computational domain with a side length of $8\kpc$ and zero-gradient boundary conditions. We choose a $64^3$ root grid and double the resolution as the distance to the BH is halved, with up to 15 refinement levels, covering a dynamic range of six orders of magnitude with a maximum resolution of $3.8\times 10^{-3}\pc$. This ensures that the void region for measuring $\MdotBH$ is resolved by at least 12 cells. Simulations are conducted in three-dimensional Cartesian coordinates to facilitate future incorporation of non-spherically symmetric effects such as turbulence and rotation. We use the MUSCL-Hancock integration scheme \cite{toro_riemann_2009}, the HLLC Riemann solver \cite{toro_riemann_2009}, and the piecewise parabolic reconstruction method \cite{woodward_numerical_1984}. Dual-energy formalism is applied to reduce truncation errors in supersonic flow.

\begin{figure*}[ht!]
\includegraphics[width=\textwidth]{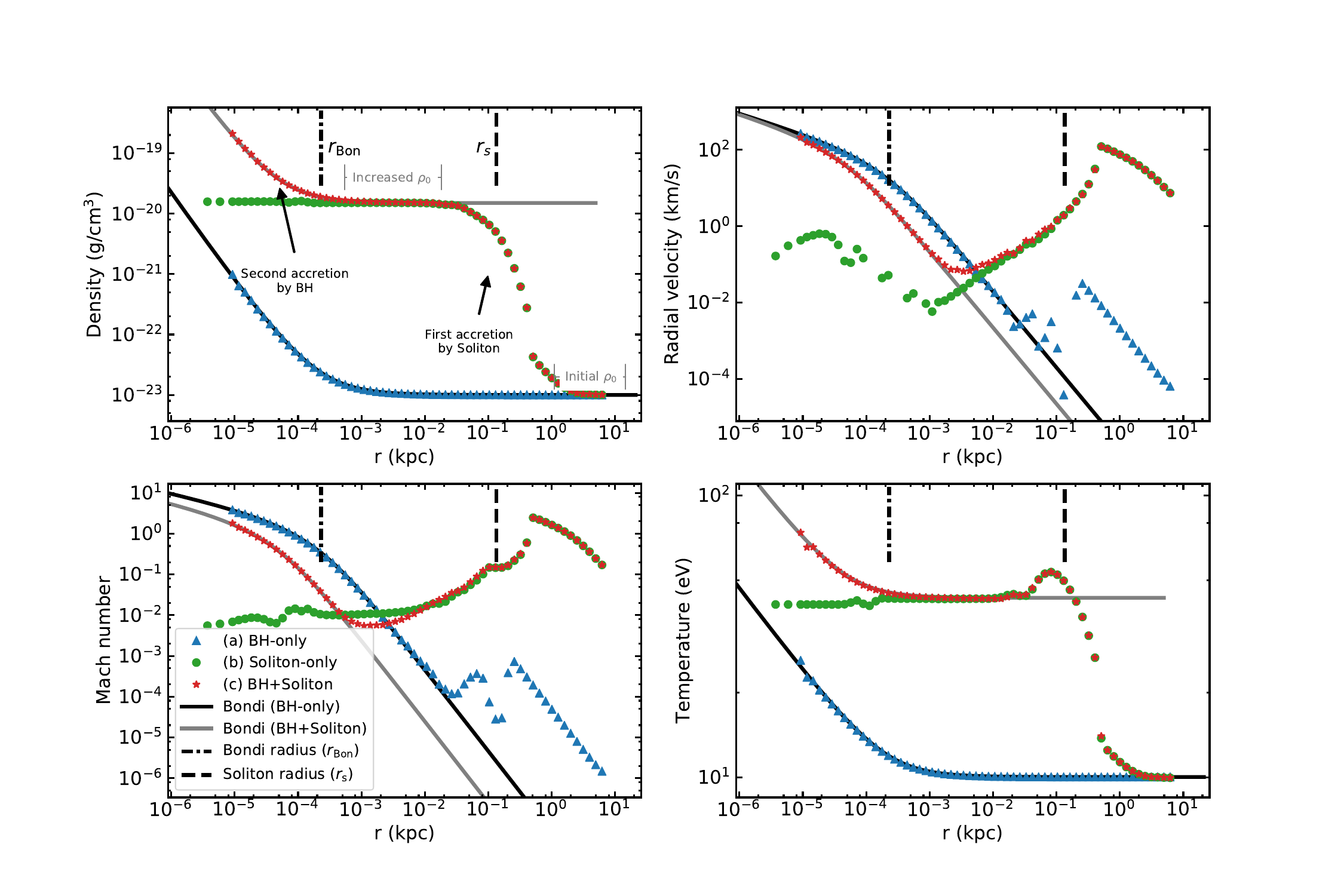}
\caption{
Gas profiles of a representative BH-soliton accretion simulation with $\gamma=1.2, T_0=10\eV$ and $m_{22}=1$. We compare three cases: (a) BH-only (b) soliton-only, and (c) both BH and soliton. Case (a) confirms that our simulation setup can reproduce the analytical Bondi solution (black solid line). Case (b) illustrates the region dominated by the soliton gravity. Case (c) presents the two-stage accretion feature of a BH-soliton system; the profile within $\rBondi$ still matches the Bondi solution (grey solid line) but with a boosting factor $\alpha \sim 125$ (see \fref{fig:gamma}) compared to the BH-only case due to the greatly increased ambient density in $\rBondi \lesssim r \lesssim r_s$ accreted by the soliton gravity. See text for details.
}
\label{fig:profile}
\end{figure*}

% BH-soliton accretion results
% ----------------------------------------------
\textit{BH-soliton accretion results}.---\fref{fig:profile} shows the gas profiles of a representative case with $\gamma=1.2$, $T_0=10\eV$, and $m_{22}=1$. We compare three scenarios: BH-only, soliton-only, and with both BH and soliton. The BH-only case confirms that our simulation setup can reproduce the analytical Bondi solution. The simulation with both BH and soliton shows a two-stage accretion attributed to  $M_s\gg \MBH$ and $r_s\gg\rBondi$. The first-stage accretion results from the soliton gravity and is stalled around $r_s$ due to its finite size. The gas density and temperature at $\rBondi \lesssim r \lesssim r_s$ increase by a factor of $\rho_{\rm inc}/ \rho_0 \sim 1000$ and $T_{\rm inc}/T_0 \sim 4$, respectively, satisfying the adiabatic condition $T \propto \rho^{\gamma-1}$. Both $\rho_{\rm inc}$ and $T_{\rm inc}$ are independent of $\MBH$, as confirmed by the consistency between the soliton-only and BH-soliton simulations at $r \gtrsim \rBondi$. 

The second-stage accretion occurs within $\rBondi$ where the BH gravity dominates over the soliton gravity. The gas profiles still satisfy the Bondi solution but with a boosting factor $\alpha \sim 125$ compared to the BH-only case due to the significantly increased ambient density from the first-stage accretion. Here we define $\alpha = \MdotBH/\MdotBondiOri$, where $\MdotBH$ is measured directly from a BH-soliton simulation and $\MdotBondiOri$ is the theoretical BH-only Bondi accretion rate from $T_0$ and $\rho_0$. Note that from $\MdotBondi \propto \rho T^{-3/2}$ and the adiabatic condition during the first-stage accretion, we can infer $\alpha = (\rho_{\rm inc}/\rho_0)^{(5-3\gamma)/2} \sim 125$, in agreement with the measured value.

% fig 2
% ----------------------------------------------

\begin{figure}[ht!]
\includegraphics[width=\columnwidth]{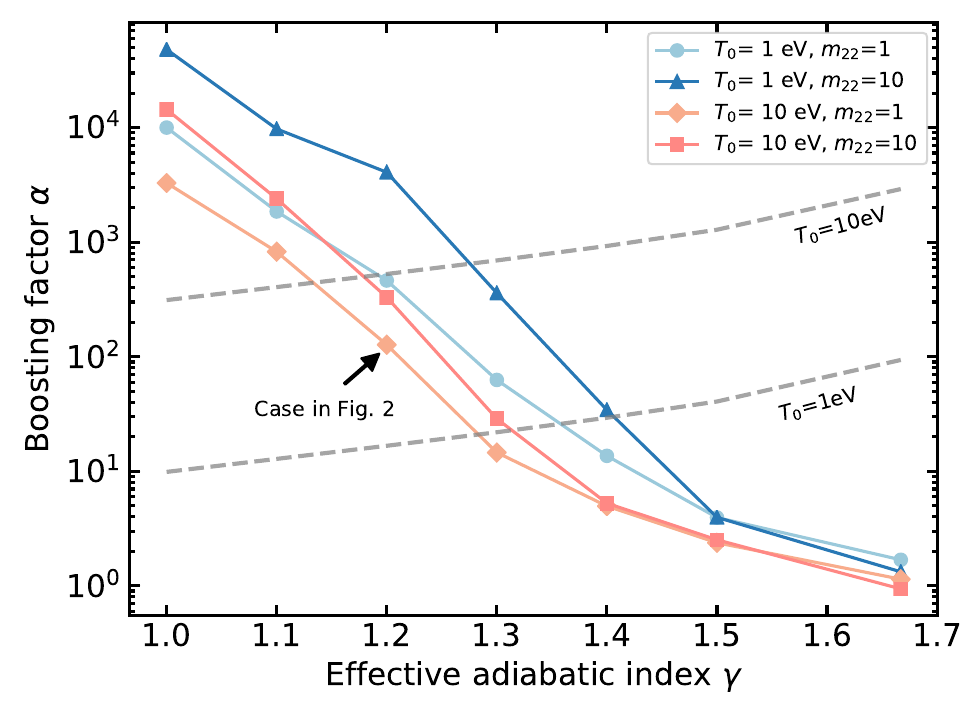}
\caption{
Strong dependence of the boosting factor $\alpha$ on the effective adiabatic index $\gamma$ for the four corner cases of ($T_0$, $m_{22}$). $\alpha$ increases with decreasing $\gamma$ and exceeds two (one) orders of magnitude when $\gamma \lesssim 1.2~(1.4)$ regardless of $T_0$ and $m_{22}$. For comparison, dashed lines show the Eddington limit with $T_0=1$ and $10\eV$. The arrow indicates the case in \fref{fig:profile}.
}
\label{fig:gamma}
\end{figure}

\fref{fig:gamma} plots the boosting factor $\alpha$ as a function of $\gamma$ for the four corner cases: ($T_0/\rm{eV}$, $m_{22}$) = (1, 1), (1, 10), (10, 1), (10, 10), revealing a strong dependence on $\gamma$ and a relatively weak dependence on $T_0$ and $m_{22}$. In general, the softer the equation of state, the larger the $\alpha$, increasing from close to unity ($\gamma = 5/3$) to three to four orders of magnitude ($\gamma = 1$) regardless of $T_0$ and $m_{22}$. This is consistent with the adiabatic prediction, $\alpha = (\rho_{\rm inc}/\rho_0)^{(5-3\gamma)/2}$, which gives $\alpha=1$ for $\gamma=5/3$ and $\alpha = \rho_{\rm inc}/\rho_0$ for $\gamma=1$. 
For comparison, we also plot the Eddington limit $\alpha=\MdotEdd/\MdotBondiOri$ with $T_0=1$ and $10\eV$, where $\dot{M}_{\rm Edd}$ is the Eddington accretion rate assuming a radiation efficiency of 10\%. Note that $\MdotBH \gtrsim \MdotEdd$ for $\gamma \lesssim 1.4$ at $T_0=1\eV$ and $\gamma \lesssim 1.2$ at $T_0=10\eV$, indicating that solitons can help approach or even exceed the Eddington limit.

% fig 3
% ----------------------------------------------

\begin{figure}[ht!]
\includegraphics[width=\columnwidth]{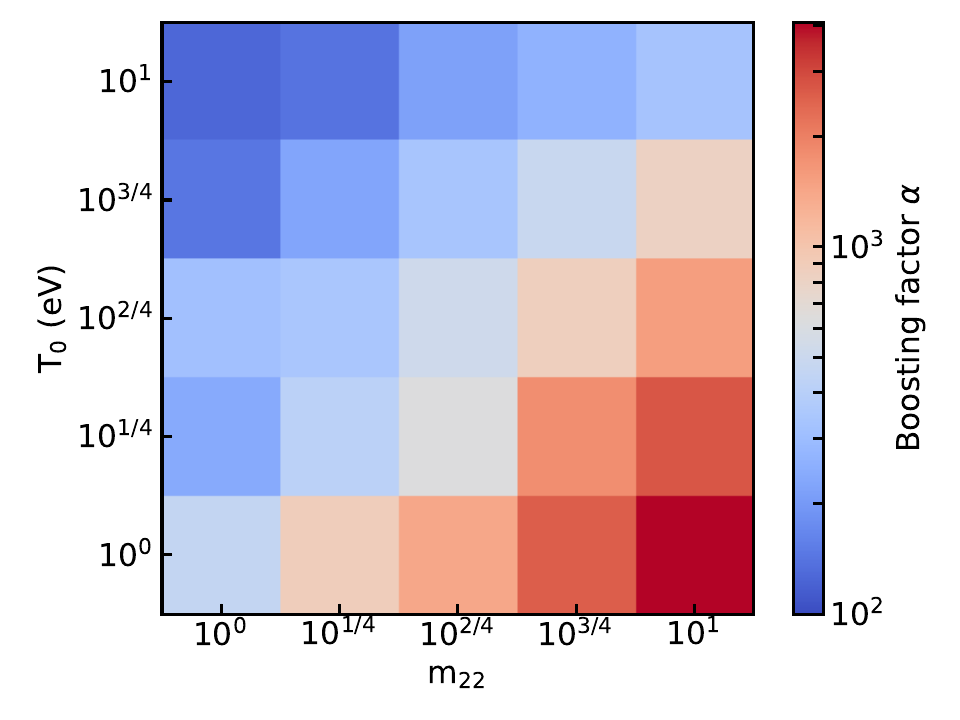}
\caption{
Boosting factor $\alpha$ as a function of $T_0$ and $m_{22}$ with $\gamma=1.2$. $\alpha$ generally exceeds two orders of magnitude and increases with increasing $m_{22}$ and decreasing $T_0$.
}
\label{fig:grid}
\end{figure}

\fref{fig:gamma} suggests that $\alpha$ increases with increasing $m_{22}$ and decreasing $T_0$ for $\gamma \lesssim 1.4$. To investigate this further, we plot in \fref{fig:grid} the boosting factor as a function of $T_0$ and $m_{22}$ for $\gamma=1.2$, showing an approximately linear dependence on $T_0^{-1}$ and $m_{22}$. This relation is attributed to the balance between gas pressure and soliton gravity in the first-stage accretion. For a given $\rho_0$, a lower $T_0$ leads to higher density compression and thus a larger $\alpha$. Similarly, for a fixed $M_h$ in Eqs. (\ref{eq:rs}) and (\ref{eq:Ms}), the gravitational acceleration near the soliton radius scales as $M_s/r_s^2 \propto (1+z)^{3/2}m_{22}$. Therefore, a larger $m_{22}$ corresponds to a stronger soliton gravity that can accrete more gas inside $r_s$ to boost $\MdotBH$. The above $(1+z)^{3/2}$ dependence also indicates that for a given $M_h$, solitons have a greater impact at higher redshifts. This can compensate for the lower $\MdotBH$ in the BH-only accretion scenario at higher redshifts when $\MBH$ is low.

% fig 4
% ----------------------------------------------

% discussion
% ----------------------------------------------
\textit{Discussion}.---The two-stage accretion occurs only when $r_s\gg\rBondi$ and $M_s\gg\MBH$. The former condition can be violated for higher $m_{22}$ or lower $T_0$, assuming a fixed $\MBH=10^5\Msun$. For example, for $m_{22}\sim 100$ and $T_0=1\eV$, $r_s\sim \rBondi$, and the accretion reverts to the single-stage process. However, $M_s\sim 10^7 \Msun \gg M_{\rm BH}$ remains ($M_s \propto m_{22}^{-1}$ from Eq. (\ref{eq:Ms})), and our simulation demonstrates a boosting factor of $\sim 1000$ for $\gamma=1.2$. For even larger particle masses in the range $10^3 \lesssim m_{22} \lesssim 10^4$, as suggested by some stringent constraints \cite{Marsh2019, Rogers2021, Dalal2022}, substantial boosting is still expected for lighter BH seeds with $\MBH \lesssim 10^5 \Msun$. As $\MBH$ grows with time, eventually $\MBH \gtrsim M_s$, and $\MdotBH$ transitions back to the conventional single-stage Bondi accretion without significant boosting.

The boosting factor $\alpha$ strongly depends on the effective adiabatic index $\gamma$, reaching $\alpha \gtrsim 10~(100)$ only when $\gamma \lesssim 1.4~(1.2)$ (\fref{fig:gamma}). What is the appropriate value of $\gamma$ in the environment of early BHs? Consider a gas halo with $T\sim1$--$10\eV$, the normalized cooling rate is $\Lambda\sim 10^{-22}$--$10^{-21}~{\rm erg\ cm^3/s}$ \cite{woodward_numerical_1984}. Assuming a gas number density of $\sim 10\ {\rm cm^{-3}}$, the cooling timescale is $\sim 0.1$--$1\kyr$, significantly shorter than the Bondi time $\tBondi$ in \tref{tab:param}. This estimate indicates efficient cooling and a low $\gamma < 5/3$ tied to a large boosting. This is further corroborated by the expectation of chaotic cold accretion frequently occurring in massive BHs and developing similarly elevated boosting factors \cite{gaspari_2017}.

To verify the robustness of our simulations, we confirm that the saturated $\alpha$ is insensitive to $\MBH$, $\rho_0$, and initial gas profiles, provided that $\MBH \ll M_s$ holds and the gas self-gravity is negligible. Additionally, $\alpha$ is independent of the radius of the central void region as long as it is within the sonic point and resolved by at least 12 cells. The artificial parameter $\beta$, used for the gradual inclusion of the soliton potential, introduces $\sim 35\%$ uncertainty in $\alpha$ but does not affect the main conclusions of this study.

Rotation, turbulence, magnetic field, AGN feedback, and radiation pressure, not currently considered in this study, can provide additional support against accretion and reduce both BH-only and BH-soliton accretion rates. Nevertheless, the boosting factor, defined as the ratio between the two rates, will likely remain substantial. Our current three-dimensional simulations make it straightforward to incorporate these additional factors, along with radiative cooling, which we will consider in future work.

We note that the boosting mechanism proposed here is likely applicable to other dark matter models that can form dense cores at high redshifts, such as self-interacting dark matter \cite{Tulin2018}.

% conclusions
% ----------------------------------------------
\textit{Conclusions}.---Soliton cores in FDM provide a potential solution to the timing problem of SMBH growth in the early universe. Our FDM cosmological zoom-in simulations with a particle mass $m_{22} \sim 1$ confirm that $\sim 10^{11}\Msun$ halos at $z \sim 7$ have well-developed solitons with masses of $\sim10^9\Msun$, comparable to the observed SMBH masses at similar epochs and significantly more massive than the BH seeds commonly employed in cosmological simulations. We then demonstrate quantitatively with hydrodynamic simulations how the additional soliton potential accretes gas to increase the ambient gas density around the BH, thereby boosting the Bondi accretion rate in its early evolution stage when the BH mass is much smaller than the soliton mass. We find that the BH accretion rate $\MdotBH$ can be boosted by two to four orders of magnitude with $m_{22}=1$--$100$, provided that radiative cooling is efficient leading to an effective $\gamma \lesssim 1.2$. Substantial boosting is also expected for $10^2 \lesssim m_{22} \lesssim 10^4$, especially for lighter BH seeds. For the same $\gamma$, $\MdotBH$ increases approximately linearly with decreasing ambient gas temperature and increasing $m_{22}$. Such boosted accretion rates due to FDM could provide a natural way for early BHs to maintain near- or super-Eddington accretion and substantially aid their rapid growth in the early universe.

% acknowledgement
% ----------------------------------------------
\begin{acknowledgments}
\textit{Acknowledgments}.---We thank Tzihong Chiueh and Tom Broadhurst for insightful discussions. We use \texttt{yt} \citep{yt} for data visualization and analysis. This research is partially supported by the National Science and Technology Council (NSTC) of Taiwan under Grant No. NSTC 111-2628-M-002-005-MY4 and the NTU Academic Research-Career Development Project under Grant No. NTU-CDP-113L7729. HYKY acknowledges support from NSTC (NSTC 112-2628-M-007-003-MY3) and Yushan Scholar Program of the Ministry of Education (MoE) of Taiwan. M.G. acknowledges support from the ERC Consolidator Grant \textit{BlackHoleWeather} (101086804).
\end{acknowledgments}

% data availability
% ----------------------------------------------
\textit{Data availability}.--- Simulation source code is available at \href{https://github.com/gamer-project/gamer}{https://github.com/gamer-project/gamer}.

% references
% ----------------------------------------------
\bibliographystyle{h-physrev}

% short names of references
\newcommand {\apjl}     {Astrophys. J. Lett. }
\newcommand {\mnras}    {Mon. Not. R. Astron. Soc. }
\newcommand {\aap}      {Astron. Astrophys. }
\newcommand {\jcap}     {J. Cosmol. Astropart. Phys. }
\newcommand {\na}       {New Astron. }
\newcommand {\araa}     {Annu. Rev. Astron. Astrophys }
\newcommand {\physrep}  {Phys. Rep. }

\bibliography{ref}

\begin{thebibliography}{100}

\bibitem{banados_800-million-solar-mass_2018}
E.~{Ba{\~n}ados} {\em et~al.},
\newblock {An 800-million-solar-mass black hole in a significantly neutral Universe at a redshift of 7.5}, \nat {\bf 553}, 473 (2018).

\bibitem{yang_poniuena_2020}
J.~Yang {\em et~al.},
\newblock P{\=o}niu{\=a}`ena: A luminous z = 7.5 quasar hosting a 1.5 billion solar mass black hole, \apjl {\bf 897}, L14 (2020).

\bibitem{larson_ceers_2023}
R.~L. {Larson} {\em et~al.},
\newblock A {CEERS} discovery of an accreting supermassive black hole 570 {Myr} after the big bang: Identifying a progenitor of massive z $>$ 6 quasars, \apjl {\bf 953}, L29 (2023).

\bibitem{kokorev_uncover_2023}
V.~{Kokorev} {\em et~al.},
\newblock {UNCOVER}: A {NIRSpec} identification of a broad-line {AGN} at z = 8.50, \apjl {\bf 957}, L7 (2023).

\bibitem{kovacs_candidate_2024}
O.~E. {Kov{\'a}cs} {\em et~al.},
\newblock A candidate supermassive black hole in a gravitationally lensed galaxy at z {\ensuremath{\approx}} 10, \apjl {\bf 965}, L21 (2024).

\bibitem{bogdan_evidence_2024}
{\'A}.~{Bogd{\'a}n} {\em et~al.},
\newblock {Evidence for heavy-seed origin of early supermassive black holes from a z {\ensuremath{\approx}} 10 X-ray quasar}, Nature Astronomy {\bf 8}, 126 (2024).

\bibitem{bosman_mature_2024}
S.~E.~I. Bosman {\em et~al.},
\newblock A mature quasar at cosmic dawn revealed by {JWST} rest-frame infrared spectroscopy, Nature Astronomy , 1 (2024).

\bibitem{suh_super-eddington-accreting_2024}
H.~Suh {\em et~al.},
\newblock A super-{Eddington}-accreting black hole {\textasciitilde}1.5 {Gyr} after the {Big} {Bang} observed with {JWST}, Nature Astronomy , 1 (2024).

\bibitem{valiante_formation_2017}
R.~Valiante, B.~Agarwal, M.~Habouzit, and E.~Pezzulli,
\newblock On the formation of the first quasars, Publications of the Astronomical Society of Australia {\bf 34}, e031 (2017).

\bibitem{woods_titans_2019}
T.~E. Woods {\em et~al.},
\newblock Titans of the early {Universe}: {The} {Prato} statement on the origin of the first supermassive black holes, Publications of the Astronomical Society of Australia {\bf 36}, e027 (2019).

\bibitem{inayoshi_assembly_2020}
K.~{Inayoshi}, E.~{Visbal}, and Z.~{Haiman},
\newblock The assembly of the first massive black holes, \araa {\bf 58}, 27 (2020).

\bibitem{zhu_formation_2022}
Q.~Zhu {\em et~al.},
\newblock The formation of the first quasars: the black hole seeds, accretion, and feedback models, \mnras {\bf 514}, 5583 (2022).

\bibitem{wise_formation_2019}
J.~H. {Wise} {\em et~al.},
\newblock {Formation of massive black holes in rapidly growing pre-galactic gas clouds}, \nat {\bf 566}, 85 (2019).

\bibitem{sassano_light_2021}
F.~Sassano {\em et~al.},
\newblock Light, medium-weight, or heavy? {The} nature of the first supermassive black hole seeds, \mnras {\bf 506}, 613 (2021).

\bibitem{mayer_direct_2024}
L.~{Mayer}, P.~R. {Capelo}, L.~{Zwick}, and T.~{Di Matteo},
\newblock Direct formation of massive black holes via dynamical collapse in metal-enriched merging galaxies at z $\sim$ 10: Fully cosmological simulations, \apj {\bf 961}, 76 (2024).

\bibitem{bromm_formation_2013}
V.~{Bromm},
\newblock {Formation of the first stars}, Reports on Progress in Physics {\bf 76}, 112901 (2013).

\bibitem{greif_numerical_2015}
T.~H. Greif,
\newblock The numerical frontier of the high-redshift {Universe}, Computational Astrophysics and Cosmology {\bf 2}, 3 (2015).

\bibitem{sugimura_birth_2020}
K.~{Sugimura}, T.~{Matsumoto}, T.~{Hosokawa}, S.~{Hirano}, and K.~{Omukai},
\newblock The birth of a massive first-star binary, \apjl {\bf 892}, L14 (2020).

\bibitem{omukai_can_2008}
K.~{Omukai}, R.~{Schneider}, and Z.~{Haiman},
\newblock Can supermassive black holes form in metal-enriched high-redshift protogalaxies?, \apj {\bf 686}, 801 (2008).

\bibitem{devecchi_formation_2009}
B.~{Devecchi} and M.~{Volonteri},
\newblock Formation of the first nuclear clusters and massive black holes at high redshift, \apj {\bf 694}, 302 (2009).

\bibitem{jeon_observability_2023}
J.~{Jeon}, B.~{Liu}, V.~{Bromm}, and S.~L. {Finkelstein},
\newblock {Observability of low-luminosity AGNs in the early Universe with JWST}, \mnras {\bf 524}, 176 (2023).

\bibitem{jeon_physical_2024}
J.~{Jeon}, V.~{Bromm}, B.~{Liu}, and S.~L. {Finkelstein},
\newblock Physical pathways for {JWST}-observed supermassive black holes in the early {Universe}, arXiv e-prints , arXiv:2402.18773 (2024).

\bibitem{bondi_spherically_1952}
H.~{Bondi},
\newblock {On spherically symmetrical accretion}, \mnras {\bf 112}, 195 (1952).

\bibitem{smole_smbh_2015}
M.~Smole, M.~Micic, and N.~Martinovic,
\newblock {SMBH} growth parameters in the early {Universe} of {Millennium} and {Millennium}-{II} simulations, \mnras {\bf 451}, 1964 (2015).

\bibitem{inayoshi_hyper-eddington_2016}
K.~{Inayoshi}, Z.~{Haiman}, and J.~P. {Ostriker},
\newblock {Hyper-Eddington accretion flows on to massive black holes}, \mnras {\bf 459}, 3738 (2016).

\bibitem{pezzulli_sustainable_2017}
E.~Pezzulli, M.~Volonteri, R.~Schneider, and R.~Valiante,
\newblock The sustainable growth of the first black holes, \mnras {\bf 471}, 589 (2017).

\bibitem{mayer_super-eddington_2019}
L.~{Mayer},
\newblock {Super-Eddington accretion; flow regimes and conditions in high-z galaxies},
\newblock in {\em Formation of the First Black Holes}, edited by M.~{Latif} and D.~{Schleicher}, pp. 195--222, World Scientific, 2019.

\bibitem{maiolino_small_2024}
R.~{Maiolino} {\em et~al.},
\newblock {A small and vigorous black hole in the early Universe}, \nat {\bf 627}, 59 (2024).

\bibitem{shi_seed_2024}
Y.~{Shi}, K.~{Kremer}, and P.~F. {Hopkins},
\newblock From seeds to supermassive black holes: Capture, growth, migration, and pairing in dense protobulge environments, \apjl {\bf 969}, L31 (2024).

\bibitem{2012MNRAS.425.2892W}
J.~S.~B. {Wyithe} and A.~{Loeb},
\newblock {Photon trapping enables super-Eddington growth of black hole seeds in galaxies at high redshift}, \mnras {\bf 425}, 2892 (2012).

\bibitem{wurster_comparative_2013}
J.~Wurster and R.~J. Thacker,
\newblock A comparative study of {AGN} feedback algorithms, \mnras {\bf 431}, 2513 (2013).

\bibitem{gaspari_2020}
M.~{Gaspari}, F.~{Tombesi}, and M.~{Cappi},
\newblock {Linking macro-, meso- and microscales in multiphase AGN feeding and feedback}, Nature Astronomy {\bf 4}, 10 (2020).

\bibitem{valentini_impact_2020}
M.~Valentini {\em et~al.},
\newblock Impact of {AGN} feedback on galaxies and their multiphase {ISM} across cosmic time, \mnras {\bf 491}, 2779 (2020).

\bibitem{yao_active_2021}
Z.~Yao, F.~Yuan, and J.~P. Ostriker,
\newblock Active galactic nucleus feedback in an elliptical galaxy with the most updated {AGN} physics: {Parameter} explorations, \mnras {\bf 501}, 398 (2021).

\bibitem{Peebles2000}
P.~J.~E. {Peebles},
\newblock Fluid dark matter, \apjl {\bf 534}, L127 (2000).

\bibitem{Guzman2000}
F.~S. {Guzm{\'a}n} and T.~{Matos},
\newblock {Scalar fields as dark matter in spiral galaxies}, Classical and Quantum Gravity {\bf 17}, L9 (2000).

\bibitem{Goodman2000}
J.~Goodman,
\newblock Repulsive dark matter, New Astronomy {\bf 5}, 103 (2000).

\bibitem{Bohmer2007}
C.~G. {B{\"o}hmer} and T.~{Harko},
\newblock {Can dark matter be a Bose Einstein condensate?}, \jcap {\bf 2007}, 025 (2007).

\bibitem{Sikivie2009}
P.~{Sikivie} and Q.~{Yang},
\newblock {Bose-Einstein} condensation of dark matter axions, \prl {\bf 103}, 111301 (2009).

\bibitem{Marsh2016}
D.~J.~E. {Marsh},
\newblock {Axion cosmology}, \physrep {\bf 643}, 1 (2016).

\bibitem{hui_ultralight_2017}
L.~Hui, J.~P. Ostriker, S.~Tremaine, and E.~Witten,
\newblock Ultralight scalars as cosmological dark matter, Phys. Rev. D {\bf 95}, 043541 (2017).

\bibitem{hui_wave_2021}
L.~{Hui},
\newblock Wave dark matter, \araa {\bf 59}, 247 (2021).

\bibitem{ChadhaDay2022}
F.~{Chadha-Day}, J.~{Ellis}, and D.~J.~E. {Marsh},
\newblock {Axion dark matter: What is it and why now?}, Science Advances {\bf 8}, eabj3618 (2022).

\bibitem{Wilczek1978}
F.~Wilczek,
\newblock Problem of strong {$P$} and {$T$} invariance in the presence of instantons, Phys. Rev. Lett. {\bf 40}, 279 (1978).

\bibitem{Arvanitaki2010}
A.~{Arvanitaki}, S.~{Dimopoulos}, S.~{Dubovsky}, N.~{Kaloper}, and J.~{March-Russell},
\newblock {String axiverse}, \prd {\bf 81}, 123530 (2010).

\bibitem{Irsic2017}
V.~Ir\ifmmode \check{s}\else \v{s}\fi{}i\ifmmode~\check{c}\else \v{c}\fi{}, M.~Viel, M.~G. Haehnelt, J.~S. Bolton, and G.~D. Becker,
\newblock First constraints on fuzzy dark matter from {Lyman}-$\ensuremath{\alpha}$ forest data and hydrodynamical simulations, Phys. Rev. Lett. {\bf 119}, 031302 (2017).

\bibitem{Rogers2021}
K.~K. {Rogers} and H.~V. {Peiris},
\newblock Strong bound on canonical ultralight axion dark matter from the {Lyman}-alpha forest, \prl {\bf 126}, 071302 (2021).

\bibitem{Davoudiasl2019}
H.~Davoudiasl and P.~B. Denton,
\newblock Ultralight boson dark matter and {Event Horizon Telescope} observations of $\mathrm{M}{87}^{*}$, Phys. Rev. Lett. {\bf 123}, 021102 (2019).

\bibitem{Brito2020}
R.~Brito, S.~Grillo, and P.~Pani,
\newblock Black hole superradiant instability from ultralight spin-2 fields, Phys. Rev. Lett. {\bf 124}, 211101 (2020).

\bibitem{Hui2023}
L.~Hui {\em et~al.},
\newblock Black hole superradiance with dark matter accretion, Phys. Rev. D {\bf 107}, 104018 (2023).

\bibitem{Chan2020}
J.~H.~H. Chan, H.-Y. Schive, S.-K. Wong, T.~Chiueh, and T.~Broadhurst,
\newblock Multiple images and flux ratio anomaly of fuzzy gravitational lenses, Phys. Rev. Lett. {\bf 125}, 111102 (2020).

\bibitem{Amruth2023}
A.~{Amruth} {\em et~al.},
\newblock {Einstein rings modulated by wavelike dark matter from anomalies in gravitationally lensed images}, Nature Astronomy {\bf 7}, 736 (2023).

\bibitem{Marsh2019}
D.~J.~E. {Marsh} and J.~C. {Niemeyer},
\newblock Strong constraints on fuzzy dark matter from ultrafaint dwarf galaxy {Eridanus II}, \prl {\bf 123}, 051103 (2019).

\bibitem{Schive2020}
H.-Y. Schive, T.~Chiueh, and T.~Broadhurst,
\newblock Soliton random walk and the cluster-stripping problem in ultralight dark matter, Phys. Rev. Lett. {\bf 124}, 201301 (2020).

\bibitem{Hayashi2021}
K.~{Hayashi}, E.~G.~M. {Ferreira}, and H.~Y.~J. {Chan},
\newblock Narrowing the mass range of fuzzy dark matter with ultrafaint dwarfs, \apjl {\bf 912}, L3 (2021).

\bibitem{Tsai2023}
Y.-D. {Tsai}, J.~{Eby}, and M.~S. {Safronova},
\newblock Direct detection of ultralight dark matter bound to the {Sun} with space quantum sensors, Nature Astronomy {\bf 7}, 113 (2023).

\bibitem{An2024}
H.~{An}, X.~{Chen}, S.~{Ge}, J.~{Liu}, and Y.~{Luo},
\newblock Searching for ultralight dark matter conversion in solar corona using {Low Frequency Array} data, Nature Communications {\bf 15}, 915 (2024).

\bibitem{Meyer2020}
Fermi-LAT Collaboration, M.~Meyer and T.~Petrushevska,
\newblock Search for axionlike-particle-induced prompt $\ensuremath{\gamma}$-ray emission from extragalactic core-collapse supernovae with the {Fermi Large Area Telescope}, Phys. Rev. Lett. {\bf 124}, 231101 (2020).

\bibitem{Ajello2016}
Fermi-LAT Collaboration, M.~Ajello {\em et~al.},
\newblock Search for spectral irregularities due to photon--axionlike-particle oscillations with the {Fermi Large Area Telescope}, Phys. Rev. Lett. {\bf 116}, 161101 (2016).

\bibitem{DeMartino2017}
I.~De~Martino {\em et~al.},
\newblock Recognizing axionic dark matter by {Compton} and de {Broglie} scale modulation of pulsar timing, Phys. Rev. Lett. {\bf 119}, 221103 (2017).

\bibitem{Blas2017}
D.~Blas, D.~L. Nacir, and S.~Sibiryakov,
\newblock Ultralight dark matter resonates with binary pulsars, Phys. Rev. Lett. {\bf 118}, 261102 (2017).

\bibitem{Smarra2023}
European Pulsar Timing Array, C.~Smarra {\em et~al.},
\newblock Second data release from the {European Pulsar Timing Array}: Challenging the ultralight dark matter paradigm, Phys. Rev. Lett. {\bf 131}, 171001 (2023).

\bibitem{Manzari2024}
C.~A. Manzari, Y.~Park, B.~R. Safdi, and I.~Savoray,
\newblock Supernova axions convert to gamma rays in magnetic fields of progenitor stars, Phys. Rev. Lett. {\bf 133}, 211002 (2024).

\bibitem{GomezBanon2024}
A.~G\'omez-Ba\~n\'on, K.~Bartnick, K.~Springmann, and J.~A. Pons,
\newblock Constraining light {QCD} axions with isolated neutron star cooling, Phys. Rev. Lett. {\bf 133}, 251002 (2024).

\bibitem{Beadle2024}
C.~Beadle, A.~Caputo, and S.~A.~R. Ellis,
\newblock Resonant conversion of wave dark matter in the ionosphere, Phys. Rev. Lett. {\bf 133}, 251001 (2024).

\bibitem{VanTilburg2015}
K.~Van~Tilburg, N.~Leefer, L.~Bougas, and D.~Budker,
\newblock Search for ultralight scalar dark matter with atomic spectroscopy, Phys. Rev. Lett. {\bf 115}, 011802 (2015).

\bibitem{Antypas2019}
D.~Antypas {\em et~al.},
\newblock Scalar dark matter in the radio-frequency band: Atomic-spectroscopy search results, Phys. Rev. Lett. {\bf 123}, 141102 (2019).

\bibitem{Stadnik2023}
Y.~V. Stadnik,
\newblock Searching for ultralight scalar dark matter with muonium and muonic atoms, Phys. Rev. Lett. {\bf 131}, 011001 (2023).

\bibitem{Bloch2023}
I.~M. Bloch {\em et~al.},
\newblock Constraints on axion-like dark matter from a {SERF} comagnetometer, Nature Communications {\bf 14}, 5784 (2023).

\bibitem{Savalle2021}
E.~Savalle {\em et~al.},
\newblock Searching for dark matter with an optical cavity and an unequal-delay interferometer, Phys. Rev. Lett. {\bf 126}, 051301 (2021).

\bibitem{Terrano2019}
W.~A. Terrano, E.~G. Adelberger, C.~A. Hagedorn, and B.~R. Heckel,
\newblock Constraints on axionlike dark matter with masses down to ${10}^{\ensuremath{-}23}\text{ }\text{ }\mathrm{eV}/{c}^{2}$, Phys. Rev. Lett. {\bf 122}, 231301 (2019).

\bibitem{Fierlinger2024}
P.~Fierlinger {\em et~al.},
\newblock Proposal for a {Ramsey} neutron-beam experiment to search for ultralight axion dark matter at the {European Spallation Source}, Phys. Rev. Lett. {\bf 133}, 181001 (2024).

\bibitem{Sikivie2021}
P.~Sikivie,
\newblock Invisible axion search methods, Rev. Mod. Phys. {\bf 93}, 015004 (2021).

\bibitem{Centers2019}
G.~P. {Centers} {\em et~al.},
\newblock {Stochastic fluctuations of bosonic dark matter}, Nature Communications {\bf 12}, 7321 (2021).

\bibitem{Branca2017}
A.~Branca {\em et~al.},
\newblock Search for an ultralight scalar dark matter candidate with the {AURIGA} detector, Phys. Rev. Lett. {\bf 118}, 021302 (2017).

\bibitem{Grote2019}
H.~Grote and Y.~V. Stadnik,
\newblock Novel signatures of dark matter in laser-interferometric gravitational-wave detectors, Phys. Rev. Res. {\bf 1}, 033187 (2019).

\bibitem{Vermeulen2021}
S.~M. {Vermeulen} {\em et~al.},
\newblock {Direct limits for scalar field dark matter from a gravitational-wave detector}, \nat {\bf 600}, 424 (2021).

\bibitem{Gottel2024}
A.~S. G\"ottel {\em et~al.},
\newblock Searching for scalar field dark matter with {LIGO}, Phys. Rev. Lett. {\bf 133}, 101001 (2024).

\bibitem{Duque2024}
F.~Duque, C.~F.~B. Macedo, R.~Vicente, and V.~Cardoso,
\newblock Extreme-mass-ratio inspirals in ultralight dark matter, Phys. Rev. Lett. {\bf 133}, 121404 (2024).

\bibitem{Lu2024}
Y.~Lu, Z.~S.~C. Picker, and A.~Kusenko,
\newblock Direct collapse supermassive black holes from relic particle decay, Phys. Rev. Lett. {\bf 133}, 091001 (2024).

\bibitem{hu_fuzzy_2000}
W.~Hu, R.~Barkana, and A.~Gruzinov,
\newblock Fuzzy cold dark matter: The wave properties of ultralight particles, Phys. Rev. Lett. {\bf 85}, 1158 (2000).

\bibitem{niemeyer_small-scale_2020}
J.~C. Niemeyer,
\newblock Small-scale structure of fuzzy and axion-like dark matter, Progress in Particle and Nuclear Physics {\bf 113}, 103787 (2020).

\bibitem{Spergel2000}
D.~N. Spergel and P.~J. Steinhardt,
\newblock Observational evidence for self-interacting cold dark matter, Phys. Rev. Lett. {\bf 84}, 3760 (2000).

\bibitem{Weinberg2015}
D.~H. {Weinberg}, J.~S. {Bullock}, F.~{Governato}, R.~{Kuzio de Naray}, and A.~H.~G. {Peter},
\newblock {Cold dark matter: Controversies on small scales}, Proceedings of the National Academy of Science {\bf 112}, 12249 (2015).

\bibitem{Kim2018}
S.~Y. {Kim}, A.~H.~G. {Peter}, and J.~R. {Hargis},
\newblock Missing satellites problem: Completeness corrections to the number of satellite galaxies in the {Milky Way} are consistent with cold dark matter predictions, \prl {\bf 121}, 211302 (2018).

\bibitem{schive_cosmic_2014}
H.-Y. {Schive}, T.~{Chiueh}, and T.~{Broadhurst},
\newblock {Cosmic structure as the quantum interference of a coherent dark wave}, Nature Physics {\bf 10}, 496 (2014).

\bibitem{schive_understanding_2014}
H.-Y. {Schive} {\em et~al.},
\newblock Understanding the core-halo relation of quantum wave dark matter from 3d simulations, \prl {\bf 113}, 261302 (2014).

\bibitem{mocz_first_2019}
P.~Mocz {\em et~al.},
\newblock First star-forming structures in fuzzy cosmic filaments, Phys. Rev. Lett. {\bf 123}, 141301 (2019).

\bibitem{chen_jeans_2017}
S.-R. {Chen}, H.-Y. {Schive}, and T.~{Chiueh},
\newblock {Jeans analysis for dwarf spheroidal galaxies in wave dark matter}, \mnras {\bf 468}, 1338 (2017).

\bibitem{Chan2022}
H.~Y.~J. {Chan}, E.~G.~M. {Ferreira}, S.~{May}, K.~{Hayashi}, and M.~{Chiba},
\newblock {The diversity of core-halo structure in the fuzzy dark matter model}, \mnras {\bf 511}, 943 (2022).

\bibitem{Bar2019}
N.~{Bar}, K.~{Blum}, J.~{Eby}, and R.~{Sato},
\newblock {Ultralight dark matter in disk galaxies}, \prd {\bf 99}, 103020 (2019).

\bibitem{Kawai2024}
H.~{Kawai}, A.~{Kamada}, K.~{Kamada}, and N.~{Yoshida},
\newblock {Modeling the core-halo mass relation in fuzzy dark matter halos}, \prd {\bf 110}, 023519 (2024).

\bibitem{Bryan1998}
G.~L. {Bryan} and M.~L. {Norman},
\newblock Statistical properties of {X}-ray clusters: Analytic and numerical comparisons, \apj {\bf 495}, 80 (1998).

\bibitem{schive_gamer-2_2018}
H.-Y. {Schive} {\em et~al.},
\newblock {GAMER-2: a GPU-accelerated adaptive mesh refinement code - accuracy, performance, and scalability}, \mnras {\bf 481}, 4815 (2018).

\bibitem{Kunkel2024}
A.~{Kunkel}, H.~Y. {Jowett Chan}, H.-Y. {Schive}, H.~{Huang}, and P.-Y. {Liao},
\newblock A hybrid scheme for fuzzy dark matter simulations combining the {Schr\"odinger} and {Hamilton-Jacobi-Madelung} equations, arXiv e-prints , arXiv:2411.17288 (2024).

\bibitem{Schwabe2022}
B.~Schwabe and J.~C. Niemeyer,
\newblock Deep zoom-in simulation of a fuzzy dark matter galactic halo, Phys. Rev. Lett. {\bf 128}, 181301 (2022).

\bibitem{AxionCAMB}
R.~Hlozek, D.~Grin, D.~J.~E. Marsh, and P.~G. Ferreira,
\newblock A search for ultralight axions using precision cosmological data, Phys. Rev. D {\bf 91}, 103512 (2015).

\bibitem{MUSIC}
O.~{Hahn} and T.~{Abel},
\newblock {Multi-scale initial conditions for cosmological simulations}, \mnras {\bf 415}, 2101 (2011).

\bibitem{Gadget2}
V.~{Springel},
\newblock {The cosmological simulation code GADGET-2}, \mnras {\bf 364}, 1105 (2005).

\bibitem{Schive2016}
H.-Y. {Schive}, T.~{Chiueh}, T.~{Broadhurst}, and K.-W. {Huang},
\newblock Contrasting galaxy formation from quantum wave dark matter, {\ensuremath{\psi}}{DM}, with {\ensuremath{\Lambda}}{CDM}, using {Planck} and {Hubble} data, \apj {\bf 818}, 89 (2016).

\bibitem{gaspari_chaotic_2013}
M.~Gaspari, M.~Ruszkowski, and S.~P. Oh,
\newblock Chaotic cold accretion on to black holes, \mnras {\bf 432}, 3401 (2013).

\bibitem{park_radiation-driven_2017}
K.~{Park}, J.~H. {Wise}, and T.~{Bogdanovi{\'c}},
\newblock Radiation-driven turbulent accretion onto massive black holes, \apj {\bf 847}, 70 (2017).

\bibitem{palous_can_2020}
J.~{Palou{\v{s}}}, S.~{Ehlerov{\'a}}, R.~{W{\"u}nsch}, and M.~R. {Morris},
\newblock {Can supernova shells feed supermassive black holes in galactic nuclei?}, \aap {\bf 644}, A72 (2020).

\bibitem{park_accelerated_2022}
K.~{Park}, G.~{Chiaki}, and J.~H. {Wise},
\newblock Accelerated growth of seed black holes by dust in the early universe, \apj {\bf 936}, 116 (2022).

\bibitem{inayoshi_rapid_2022}
K.~{Inayoshi} {\em et~al.},
\newblock Rapid growth of seed black holes during early bulge formation, \apj {\bf 927}, 237 (2022).

\bibitem{toro_riemann_2009}
E.~Toro,
\newblock {\em Riemann Solvers and Numerical Methods for Fluid Dynamics: A Practical Introduction}, 3rd ed. (Berlin: Springer, 2009), .

\bibitem{woodward_numerical_1984}
K.~M. {Schure}, D.~{Kosenko}, J.~S. {Kaastra}, R.~{Keppens}, and J.~{Vink},
\newblock {A new radiative cooling curve based on an up-to-date plasma emission code}, \aap {\bf 508}, 751 (2009).

\bibitem{Dalal2022}
N.~{Dalal} and A.~{Kravtsov},
\newblock {Excluding fuzzy dark matter with sizes and stellar kinematics of ultrafaint dwarf galaxies}, \prd {\bf 106}, 063517 (2022).

\bibitem{gaspari_2017}
M.~{Gaspari}, P.~{Temi}, and F.~{Brighenti},
\newblock {Raining on black holes and massive galaxies: the top-down multiphase condensation model}, \mnras {\bf 466}, 677 (2017).

\bibitem{Tulin2018}
S.~{Tulin} and H.-B. {Yu},
\newblock {Dark matter self-interactions and small scale structure}, \physrep {\bf 730}, 1 (2018).

\bibitem{yt}
M.~J. {Turk} {\em et~al.},
\newblock yt: A multi-code analysis toolkit for astrophysical simulation data, Astrophys. J. Suppl. {\bf 192}, 9 (2011).

\end{thebibliography}

\end{document}